\documentclass[traditabstract]{aa} 
\usepackage{graphicx}
\usepackage{txfonts}
\begin{document}
\renewcommand{\textfraction}{0.0}
\renewcommand{\bottomfraction}{1.0}
\renewcommand{\dblfloatpagefraction}{1.00}

   \title{Outflow and accretion detections in the young stellar object IRAS 04579+4703}
  \author{Jin-Long Xu,
          \inst{1,2}
          Jun-Jie Wang\inst{1}
          \ and Sheng-Li Qin\inst{3}
          }
   \institute{ National Astronomical Observatories, Chinese Academy of Sciences,
             Beijing 100012, China \\
         \email{xujl@bao.ac.cn}
         \and
          Graduate University of the Chinese Academy of Sciences, Beijing, 100080, China\\
        \and
             I. Physikalisches Institut, Universit\"{a}t zu K\"{o}ln, Z\"{u}lpicher Str. 77, 50937 K\"{o}ln, Germany\\
        }
\authorrunning{J.-L. Xu et al.}
\titlerunning{Outflow and accretion  in  IRAS 04579+4703}
\abstract {We present  Submillimeter Array observations of the young
stellar object IRAS 04579+4703 in the 1.3 mm continuum and in the
$\rm ^{12}CO (2-1)$, $\rm ^{13}CO(2-1)$ and $\rm C^{18}O(2-1)$
lines. The 1.3 mm continuum image reveals a flattened structure with
a mass of 13 $M_{\odot}$. The $\rm ^{12}CO (2-1)$ line map and
position-velocity (PV) diagram, together with the broad wing (full
width = 30 km $\rm s^{-1}$) of $\rm ^{12}CO (2-1)$ line, clearly
show that  there is an outflow motion, which originates from an
embedded massive YSO in this region. The lengths of the blue-shifted
and red-shifted  lobes are 0.14 pc and 0.13 pc respectively. The
total gas mass, average dynamical timescale and mass entrainment
rate of the outflow are 1.8 $M_{\odot}$,  1.7 $\times$ $10^{4}$ yr
and 1.1 $\times$ $10^{-4}$ $M_{\odot}\ \rm yr^{-1}$, respectively.
The flattened  morphology of the continuum source perpendicular to
the outflow direction,  and the velocity gradient seen in the
spectra of $\rm C^{18}O (2-1)$ taken from different locations along
the major axis of the continuum source, suggest the presence of an
accretion disk in this region.}


   \maketitle
%
%

\section{Introduction}
Observational evidences suggest that massive stars may be formed
through the accretion-disk-outflow processes similar to low-mass
stars (e.g, Shu et al. \cite{Shu87}; Cesaroni et al. 1999; Shepherd
et al. 2001;Patel et al. \cite{Patel}; Jiang et al. \cite{jiang};
Zinnecker \& Yorke \cite{zin}). Searching for  collapse, accretion
and outflow with   interferometers  in  massive star formation
regions has boomed (e.g, Beuther e al. \cite{Beuther04}; Qin et al.
\cite{qin08}; Furuya et al. \cite{Furuya2011}),  but the limited
observations  cannot provide detailed information for our
understanding of massive star formation. The major difficulty is due
to  the large distances ($\geq1$ kpc), clustered formation
environments and shorter evolutionary timescale of massive stars.

IRAS 04579+4703 is a young  high mass stellar object (YSO). Molinari
et al. (\cite{Mol96}) carried out the $\rm NH_{3}$ observation
toward IRAS 04579+4703, and obtained a distance of 2.47 kpc from the
sun and a bolometric luminosity of 3.91$\times$10$^{3}L_{\odot}$.
However, Molinari et al. (\cite{Mol98}) failed to observe any 6 cm
emission from IRAS 04579+4703. S\'{a}nchez-Monge et al.
(\cite{San08}) performed higher spatial resolution continuum
observations at 1.2 mm, 7 mm, 1.3 cm and 3.6 cm toward IRAS
04579+4703.  Their observations showed that the 7 mm emission,
similar to the 1.3 cm and 3.6 cm emission in morphology, is
elongated in the southwest-northeast direction. The location of
their 1.2 mm continuum peak agrees well with that of the source
detected at cm wavelengths as well as an $\rm H_{2}O $ maser (Palla
et al. \cite{Palla91} \& Migenes et al. \cite{Mig99}), indicating
that the emissions only concentrate on a small and compact region
and the IRAS 04579+4703 is at an early evolutionary stage. Using the
SED fit, S\'{a}nchez-Monge et al. (\cite{San08}) derived {\bf a}
dust temperature of about 30 K and obtained a gas mass of 23
$M_{\odot}$.  The line $\rm ^{12}CO (2-1)$ observations of Zhang et
al. (\cite{zha05}) did not reveal molecular outflow from IRAS
04579+4703, while Wouterloot \& Brand (\cite{Wou89})  observed the
$\rm ^{12}CO (1-0)$ emission with a linewidth of 6.1 km $\rm
s^{-1}$. Additionally, Varricatt et al. (\cite{Var10}) detected
jet-like $\rm H_{2}$ knots in the northwest-southeast direction.
However, these observations can not provide  sufficient evidences
for the outflow.  Higher spatial resolution observations of the
molecular lines and continuum in submillimeter/millimeter waveband
are needed to reveal the detailed source structure and kinematics in
this region.

 In this paper we present Submillmeter Array (SMA) observations
of the young stellar object IRAS 04579+4703 obtained in the 1.3 mm
continuum and in the $\rm ^{12}CO (2-1)$, $\rm ^{13}CO(2-1)$, and
$\rm C^{18}O(2-1)$. Our observations at high angular resolution
reveal bipolar outflow motion and gas accretion for IRAS 04579+4703.

\section{OBSERVATIONS AND DATA REDUCTION}

The observations toward IRAS 04579+4703 were performed with the SMA
on 2008 March 21, at 220 (lower sideband) and 230 GHz (upper
sideband)\footnote{The data are publicly available on the website
http://www.cfa.harvard.edu/rtdc/data/search.html}. The two sidebands
of the SMA covered frequency ranges of 219.4--221.4 GHz and
229.4--231.4 GHz, respectively. The projected baselines ranged from
7 k$\lambda$ to 100 k$\lambda$. The phase  track center was
R.A.(J2000.0)= $\rm 05 ^{h}01^{m}39^{s}.92$ and  Dec.(J2000.0) =
$\rm 47 ^{\circ}07^{\prime}21^{\prime\prime}.10$. The typical system
temperature was 186 K. The spectral resolution is 0.406 MHz,
corresponding to a velocity resolution of 0.5 km $\rm s^{-1}$.  The
bright quasar 3C279 was used for bandpass calibration, while
absolute flux density scales were determined from observations of
Titan. QSO 0533+483 and QSO 0359+509 were observed for  antenna gain
corrections. The calibration and imaging were performed in Miriad.
A continuum image was constructed from the line-free channels. The
spectral cubes were constructed using the continuum-subtracted
spectral channels. Self-calibration was performed to the continuum
data. The gain solutions from the continuum were applied to the line
data. The synthesized beam size of the continuum  was approximately
$3^{\prime\prime}.49$ $\times$ $2^{\prime\prime}.63$ with a P.A.
=-$58^{\circ}.9$.  Based on the phase monitoring of QSO 0533+483,
the absolution position accuracy (pointing error) is estimated to be
$\sim$0.1$^{\prime\prime}$ during the observations.

\begin{figure}[]
\vspace{0mm}
\includegraphics[angle=270,scale=.32]{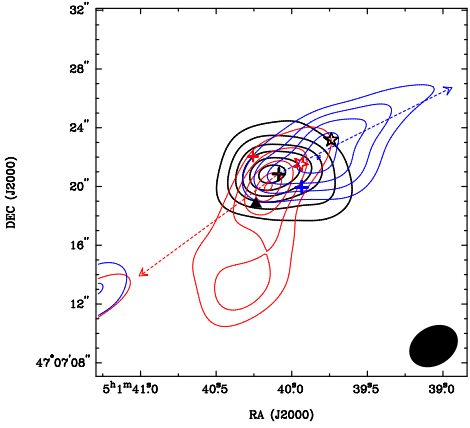}
\vspace{-8mm}\caption{1.3 mm continuum emission (black thicker
contours) overlaid with the velocity integrated intensity map of
$\rm ^{12}CO(2-1)$ outflow (red and blue contours). The black
contour levels are at  3, 6, 9, 12, 15, 18 and 21 $\sigma$ (1
$\sigma$ is 0.002 $\rm Jy\ beam^{-1}$).  The red and blue contour
levels are  35, 50,..., 95 $\%$ of the peak value. The synthesized
beam is $3^{\prime \prime}.49\times2^{\prime \prime}.63$, P.A.=
-58.9$^{\circ}$ (shown in the lower right corner). The black plus
symbol marks the peak position of the continuum source.  The blue
and red plus symbols represent the positions of the extracted
spectra in Fig.4, respectively. $\rm H_{2}O$ maser is shown with the
filled triangle. The black and red open star symbols indicate the
location of the infrared source IRAS 04579+4703 and a near-IR source
detected by Varricatt et al. (\cite{Var10}), respectively. The blue
and red dashed lines mark the direction of the NW and SE lobes of
the $\rm H_{2}$  knots (Varricatt et al. \cite{Var10}),
respectively.}
\end{figure}

\section{RESULTS}
\subsection{Continuum Emission at 1.3 mm}
In Figure 1, the 1.3 mm continuum  image observed with the SMA shows
 a flattened source structure. Using a two-dimensional Gaussian
fit to the  continuum, we  obtained a total integrated flux density
of 0.11 Jy, a deconvolved source size  of 4.7$^{\prime\prime}$
$\times$ 2.9$^{\prime\prime}$ (P.A. = 83.1$^{\circ}$), and a peak
position of R.A.(J2000) = 05$\rm ^{h}$01$\rm ^{m}$40.$\rm ^{s}$086,
 Dec.(J2000) =
+47$^{\circ}$07$^{\prime}$20.$^{\prime\prime}$90. The positional
uncertainty due to noise can be estimated by $\triangle \rm\theta$ =
0.45 $\frac{\theta_{\rm FWHM}}{\rm S/N}$, where $\theta_{\rm FWHM}$
is the angular resolution and S/N is the signal-to-noise ratio (Reid
et al. \cite{Reid88}; Chen et al. \cite{Chen06}; Weintroub et al.
\cite{Wei08}; Qin et al. \cite{qin10}). The estimated positional
uncertainty is approximately $0.^{\prime\prime}14$. If we also
consider the uncertainty from the Gaussian fit to the position of
the continuum peak  and the pointing error, then $\Delta$R.A. =
$\pm$0.24$^{\prime\prime}$ and $\Delta$Dec. =
$\pm$0.21$^{\prime\prime}$. The  1.3 mm continuum emission is
associated with an $\rm H_{2}O$ maser (Migenes et al. \cite{Mig99})
as well as a near-IR source detected by Varricatt et al.
(\cite{Var10}). In addition, the peak of the  1.3 mm continuum
agrees well with those of the 1.3 cm, 3.6 cm and 7 mm emission
within the positional uncertainty (S\'{a}nchez-Monge et al.
\cite{San08}). S\'{a}nchez-Monge et al. (\cite{San08})  argued that
these radio emission could be from either radio jets or hypercompart
$\rm H \scriptstyle II$  regions.  The rising spectral index
(S\'{a}nchez-Monge et al. \cite{San08}), association with an $\rm
H_{2}O $ maser, and compact source structure suggest a hypercompact
$\rm H \scriptstyle II$ region in this region (Sewilo et al. 2004).
Therefore the 1.3 mm continuum emission may comprise free-free
emission from the hypercompart $\rm H \scriptstyle II$ region and
 the thermal emission from the warm dust from its surroundings
(Keto et al. \cite{Keto} \& Beltr\'{a}n et al. \cite{Beltran}). The
emission from the free-free component of the continuum flux in our
source can be estimated by $S_{\nu} \propto \nu^{\alpha}$, where the
free-free spectral index $\alpha$ is 1.1 from 3.6 to 1.3 cm
continuum emission. The contribution of the free-free emission to
the 1.3 mm continuum  is $\sim$ 4.4 mJy (much lower than our total
integrated flux 0.11 Jy at 1.3 mm), and then can safely be ignored.

In order to determine the total gas mass of the continuum source, we
make the optically thin approximation for the dust continuum
emission. The total  gas mass of  the continuum source can be
estimated by $M_{\rm core}=S_{v}D^{2}/ \kappa_{\nu}RB_{\nu}(T_{d})$,
where $S_{v}$ is the flux at the frequency $\nu$ and $D$ is the
distance to the source. We assume $\kappa_{\nu}$ =
$0.05(\nu/230)^{\beta}$ m$^{2}$ kg$^{-1}$ for the opacity (Testi \&
Sargent \cite{Tes98}), where we adopt $\beta$ = 2 (Drain \& Lee
\cite{Draine84}) and a dust-to-gas mass ratio $R$ = 0.01 (Lis et al.
\cite{Lis91}). $B_{v}(T_{\rm d})$ is the Planck function for a dust
temperature $T_{d}$ at frequency $\nu$. Using a dust temperature of
$\sim$ 30 K (S\'{a}nchez-Monge et al. \cite{San08}), we derive total
gas mass = 13 $ M_{\odot}$. Because the extended envelope can be
filtered by the  interferometers (missing short spacing data), the
mass of the gas derived from our data is smaller than 23 $
M_{\odot}$ from the single dish observations (S\'{a}nchez-Monge et
al. \cite{San08}).

\begin{table*}
\tabcolsep 1.7mm\caption{The physical parameters of the outflow. }
\vspace{-3mm}
\begin{tabular}{lcccccccccccc}
\hline\hline
  Wing & Velocity interval& Deconvolved sizes &$T_{\rm mb}$ & $T_{\rm ex}$ &&$\tau$& $N_{\rm H_{2}}$& $M_{\rm out}$  & $t_{d}$ &$\dot{M}_{\rm out}$         \\
       & km $\rm s^{-1}$ & arcsec $\times$ arcsec & K  & K & & & ($\times10^{21}\rm cm^{-2}$)&($\rm M_{\odot}$)& ($\rm \times10^{4}yr$) &($\times10^{-5}$$\rm M_{\odot}\ yr^{-1}$)        \\
  \hline\noalign{\smallskip}
   Blue & -24.2 to -20.0& 9.9$\times$4.3(P.A.=-60.5$^{\circ}$)&10.0&15.2 & &3.7& 3.1 & 0.4 & 1.9 & 2.1    \\  
   Red  & -13.0 to -8.5& 13.3$\times$5.2(P.A.=-22.5$^{\circ}$)&7.8 &12.8 & &3.3& 6.5 & 1.4 & 1.5 & 9.3   \\
\noalign{\smallskip}\hline
\end{tabular}
\end{table*}

\subsection{Molecular Line Emission}
 Figure 2 shows the spectra of $\rm ^{12}CO (2-1)$, $\rm
^{13}CO(2-1)$ and $\rm C^{18}O(2-1)$  emission at the peak position
of the continuum image. The $\rm ^{12}CO (2-1)$ and $\rm
^{13}CO(2-1)$ lines present asymmetric profile with double peaks
which can be caused by large optical depth (Andr\'{e} et al.
\cite{andre} \& Belloche et al. \cite{Belloche02}).  The $\rm
C^{18}O(2-1)$ line with a single emission peak can be considered as
optically thin; Hence it can be used for determination of the
systemic velocity. We measure a systemic velocity of $\sim$ $-$17.0
km $\rm s^{-1}$ from this line. The velocities of the blue and red
emission peaks of $\rm ^{12}CO (2-1)$ are $-$20 km s$^{-1}$ and
$-$13 km s$^{-1}$. For $\rm ^{13}CO (2-1)$, the velocities of the
blue and red emission peaks are $-$19 km s$^{-1}$ and $-$16 km
s$^{-1}$.

The $\rm ^{12}CO (2-1)$ and $\rm ^{13}CO(2-1)$ spectra show $``$blue
profile$"$ signature (Wu et al. \cite{Wu07}), and the blue peaks of
both lines are stronger than their red peaks with an absorption dip
near the systemic velocity, which can be produced by infall motion
or accretion. In addition, the $\rm ^{12}CO (2-1)$ spectrum show an
unusual full width of $\sim$30 km $\rm s^{-1}$, from $-$38 to $-$8
km $\rm s^{-1}$. Such a broad wing is a strong indication of outflow
motion.

\begin{figure}[]
\vspace{-16mm}
\includegraphics[angle=180,scale=.49]{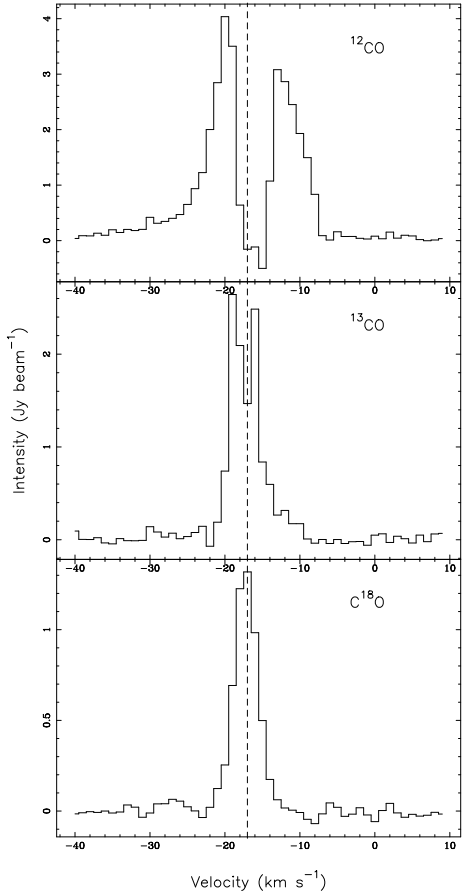}
\vspace{2mm}\caption{Spectral profiles observed at the central
position of IRAS 04579+4703 in the optically thick $\rm ^{12}CO
(2-1)$, $\rm ^{13}CO(2-1)$ lines, and the optically thin $\rm
C^{18}O(2-1)$ line. The dotted line in the spectra marks the cloud
systemic velocity.}
\end{figure}

To determine the velocity components and spatial extension of the
outflow, we have made position-velocity (PV) diagram  along the cuts
at various position angles, and found that the redshifted and
blueshifted components are  the strongest at a position angle of
$-40^{\circ}$ (40$^{\circ}$ west of north) as shown in Figure 3. The
PV diagram in Fig. 3 clearly shows bipolar components. The
blueshifted and redshifted components have obvious velocity
gradients from $-$24.2 to $-$20 km $\rm s^{-1}$ and $-$13 to $-$8.5
km $\rm s^{-1}$ respectively (see  the  blue and red vertical dashed
lines in Fig. 3). The distributions of redshifted and blueshifted
velocity components in Fig.3 provide us further evidence for the
bipolar outflow in this region (Yamashita et al.
\cite{Yamashita89}).

Using the velocity ranges determined from the PV diagram,  we made
the velocity integrated intensity maps superimposed on the 1.3 mm
continuum as shown in Fig. 1. The blueshifted and redshifted
components are presented as blue and red contours. The map clearly
shows northwest-southeast bipolar components centered at the peak
position of the 1.3 mm continuum. The blueshifted and redshifted
lobes are not  in a straight line. The outflow is associated with
infrared source IRAS 04579+4703 and a near-IR source.

\begin{figure}[]
\vspace{-6mm}
\includegraphics[angle=270,scale=.31]{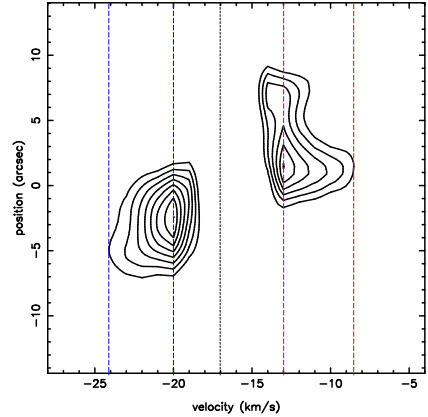}
\vspace{-1mm}\caption{ The position-velocity plot of the $\rm
^{12}CO(2-1)$ emission at a position angle of $-40^{\circ}$. The
contour levels are 1.4, 1.8, 2.3, 2.7, 3.2, 3.6, 4.1 and 4.5 Jy
beam$^{-1}$. 1 $\sigma$ is 0.7 Jy beam$^{-1}$. The black vertical
dashed line marks the cloud systemic velocity. The blue and red
vertical dashed lines mark the velocity ranges of the blueshifted
and redshifted emission, respectively.}
\end{figure}

\section{DISCUSSION}
\subsection{Outflow Kinematics and Driving Source}
The  velocity integrated intensity map, PV diagram and the broad
wing (full width = 30 km $\rm s^{-1}$) of the CO $\rm ^{12}CO (2-1)$
line clearly show a bipolar outflow in  IRAS 04579+4703 region. The
blueshifted and redshifted lobes do not lie in a straight line, and
the extension is similar to that of  $\rm H_{2}$ knots (Varricatt et
al. \cite{Var10}), indicating that the H2 and CO most likely
represent the same outflow,  and that the $\rm ^{12}CO (2-1)$
outflow gas probably is entrained by the $\rm H_{2}$ jet-like
outflow. The outflow is associated with the infrared source IRAS
04579+4703 and a  near-IR source. Varricatt et al. (\cite{Var10})
suggested that the near-IR source may be the near-IR counterpart of
IRAS 04579+4703 with the strong infrared excess. The position of the
near-IR source agrees well with the peaks of the 1.3 mm, 7 mm, 1.3
cm and 3.6 cm emission within the positional uncertainty. H2 and
Br$\gamma$ emission are very close to the near-IR source (Ishii et
al. \cite{Ishii}  \& Varricatt et al. \cite{Var10}). Together, we
conclude that the near-IR  source may be an embedded massive YSO,
which is likely to drive the bipolar outflow.

Under conditions of local thermodynamical equilibrium (LTE), $\rm
^{12}CO$(2-1) being optically thick, and the completely filling beam
with the source (the filling factor $f=1$), the column density of
the outflow is estimated by using the equation (eq. 1) following
Garden et al. (\cite{Garden91}) with a relation $N_{\rm H_{2}}$
$\approx$ $10^{4}N_{\rm ^{12}CO}$( Dickman \cite{Dickman78}):
\begin{equation}
\mathit{N_{\rm H_{2}}}=1.08\times10^{17}\frac{(T_{\rm
ex}+0.92)}{\exp(-16.62/T_{\rm ex})}\int T_{\rm mb}\times\frac{\tau
dv}{[1-\exp(-\tau)]}\rm cm^{-2},
\end{equation}

\indent where $T_{\rm ex}$ is the excitation temperature, $T_{\rm
mb}$ is the corrected main beam temperature, and $\tau$ is the
optical depth for  the $\rm ^{12}CO$(2-1) line.  We assume the solar
abundance ratio [$^{12}$CO]/[C$^{18}$O]=$\tau(^{12}$CO)/$\tau(\rm
C^{18}$O)=490 (Garden et al. \cite{Garden91}). $T_{\rm ex}$ is
estimated following the equation $T_{\rm ex}=11.1/{\ln[1+1/(T_{\rm
mb}/11.1+0.02)]}$ (Garden et al. \cite{Garden91}). The red and blue
lobes of the outflow peak near the continuum, then we assume that
the $^{12}$CO and C$^{18}$O trace same cloud components in the peak
positions of the two lobes for the calculation, the assumption is
reasonable since both  $^{12}$CO$(2-1)$ and C$^{18}$O$(2-1)$ are low
energy transition with similar upper level energy. Under assumption
of C$^{18}$O being optically thin,  we calculate $\tau$ according to
$\tau=\rm-490ln[1-(e^{11.1/\it T_{\rm ex}}-1) \it $$T^{18}_{\rm
mb}$$/\rm 11.1]$ (Garden et al. \cite{Garden91}). Under the
Rayleigh-Jeans approximation, 1 Jy beam$^{-1}$ in our SMA
observations corresponds to a brightness temperature of 2.52 K.
Furthermore, the mass of the outflow is given by $M_{\rm
out}=mN_{\rm H_{2}}S/(2.0\times10^{33})$ $\rm M_{\odot}$, where $m$
is 1.36 times the $\rm H_{2}$ mass (Garden et al. \cite{Garden91})
and $S$ is the size of each lobe. $S$ is obtained from the
deconvolved lobe size (In Table 1). A dynamic time scale can be
determined by $t_{d}=r/v$, where $v$ is the maximum flow velocity
relative to the cloud systemic velocity , and $r$ is the length of
the begin-to-end flow extension for each lobe. The lengths of the
blueshifted and redshifted lobe are 0.14 pc and 0.13 pc obtained
from Fig.1, respectively, which is only a lower limit due to
projection on the sky plane. The mass entrainment rate of the
outflow is calculated using $\dot{M}_{\rm out}=M_{\rm out}/(t_{\rm
d})$. The physical parameters and the calculated results of the
outflow are summarized in Table 1.  The parameter difference of the
two lobes reflects that the ambient material is not uniform and then
the entrained masses and dynamics are different.  The outflow has
the total mass of 1.8 $M_{\odot}$ and the mass entrainment rate of
1.1 $\times$ $10^{-4}$ $M_{\odot}\ yr^{-1}$. The average dynamical
timescale is about 1.7 $\times$ $10^{4}$ yr.

\subsection{Accretion}
Our interferometric continuum  image toward IRAS 04579+4703 reveals
 a flattened structure, perpendicular to the direction of the
outflow detected in $\rm ^{12}CO (2-1)$ and $\rm H_{2}$, which is
consistent with accretion disk scenario observed in the other high
mass star formation regions (Cesaroni et al. 1999; Shepherd et al.
2001; Patel et al. 2005). The $\rm C^{18}O(2-1)$ spectra from the
two positions (the red and blue plus symbols in Fig. 1)
perpendicular to the direction of the outflow are shown in Figure 4.
The velocity gradient is seen from the $\rm C^{18}O(2-1)$ spectra,
suggesting that the disk is rotationally and gravitationally bound
by the central YSO.  Assuming the rotation is Keplerican, the
dynamical mass (binding mass) is estimated by $M = \delta
v^{2}r/G\rm sin^{2}$$(i)$, where $r$ is the half spatial separation
of the emission, $\delta v$ is the difference between the peak
velocity of the emission and the systemic velocity, and $i$ is the
unknown inclination angle between the disk plane and the plane of
the sky ($i$ = 90$^\circ$ for an edge-on system). Adopting the
values $r$ $\sim$2$^{\prime\prime}$ and $\delta v$ $\sim$ 1 km
s$^{-1}$ shown in Figures 1 and 4, the obtained dynamical mass is
$\sim$6/[sin$^{2}(i)$]$M_{\odot}$.  The dynamical mass is estimated
to be 12 $M_{\odot}$ if  $i$ = 45$^{\circ}$, which is a lower limit
since that we estimated the binding mass from the kinematics inside
2$^{\prime\prime}$ due to our low spatial resolution observations.

\begin{figure}[]
\vspace{-7mm}
\includegraphics[angle=270,scale=.34]{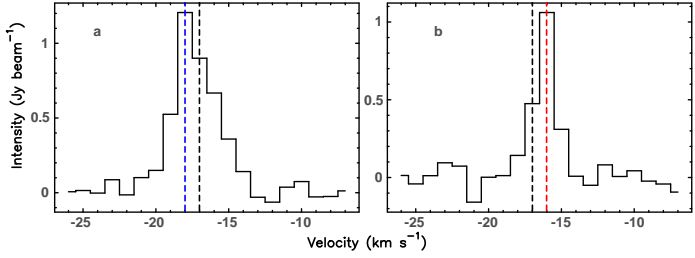}
\vspace{-23mm}\caption{$\rm C^{18}O(2-1)$ spectra. In a and b
panels, the molecular spectra are extracted from the positions of
the blue and red plus symbols marked in Fig. 1, respectively, which
are separated by 4$^{\prime\prime}$. The black vertical dashed line
marks the cloud systemic velocity. The blue and red vertical dashed
lines mark the peak velocities, respectively.} \vspace{-4mm}
\end{figure}

The spectroscopic observations of H$_2$ and Br$\gamma$ (Ishii et al.
\cite{Ishii}) also suggested the presence of an accretion disk, and
this source is probably accreting. If we assume that the central
star is a massive protostar and not in the phase of zero-age
main-sequence (ZAMS) star, then the bolometric luminosity
(3.91$\times$10$^{3}L_{\odot}$) is equal to accretion luminosity.
According to Molinari et al. (\cite{Mol981}), the accretion rate is
given by
$\dot{M}=6.22\times10^{-9}(L/L_{\odot})^{1.70}(M/M_{\odot})^{-1.24}$
. Assuming the dynamical mass of 12 $M_{\odot}$ is equal to the
protostar mass, the expected accretion rate is 3.7$\times10^{-4}$
$M_{\odot} \rm yr^{-1}$.

The $``$blue profile$"$ of the $\rm ^{12}CO (2-1)$ line provide a
further evidence for infall motion or accretion as discussed in
section 3.2.  The mass accretion rate is estimated by $\dot{M}=4 \pi
\it R^{\rm 2}Vmn$ (Myers et al. \cite{Mye96}), where $m$ is 1.36
times the $\rm H_{2}$ mass. If cloud core is approximately spherical
in shape, the mean number density is $n$$=1.62\times10^{-19}N_{\rm
H_{2}}/L$,  where $L$ is the cloud core diameter in parsecs, which
is 0.056 pc (the major axis of the deconvolved size of the
continuum). $N(\rm H_{2})$ is the column density of molecular
hydrogen obtained from the equation (eq. 3) of Xu et al.
(\cite{xu10}) in C$^{18}$O line. The velocity difference ($V$) of
2.0 km s$^{-1}$ between the systemic velocity (-17.0 km s$^{-1}$)
and the velocity of the redshifted absorbing dip (-15.0 km s$^{-1}$)
and a source size ($R$) of 0.028 pc, then we obtain a mass accretion
rate of $\sim$ 2.3$\times10^{-4}$ $M_{\odot} \rm yr^{-1}$ which is
roughly consistent with the value derived from bolometric
luminosity, further suggesting the presence of an accretion disk in
this region, but the observations at the current spatial resolution
are not adequate to distinguish the disk accretion and  the
observations of more molecular tracer with higher spatial
resolutions are needed to resolve detailed kinematics in this
region.

\begin{acknowledgements}  We thank the anonymous referee for his/her constructive comments and
suggestions that greatly improved the content and presentation of
this paper. Jin-Long Xu's research is in part supported by 2011
Ministry of Education doctoral academic prize.
\end{acknowledgements}

\end{document}